\documentclass[11pt]{iopart}

\expandafter\let\csname equation*\endcsname\relax
\expandafter\let\csname endequation*\endcsname\relax 

\usepackage[english]{babel}
\usepackage[utf8]{inputenc}
\usepackage{amsmath}
\usepackage{graphicx}
\usepackage[colorinlistoftodos]{todonotes}
\usepackage{hyperref}
\usepackage{epsfig}
\usepackage{subcaption}
\usepackage{booktabs}
\usepackage{sidecap}
\begin{document}

\title{Emergence of Cooperative Long-term Market Loyalty in Double Auction Markets}

\author{Aleksandra Alori\'c$^\dag$, Peter Sollich$^\dag$, Peter McBurney$^\ddag$, Tobias Galla$\S$}
\address{$\dag$ ~ Department of Mathematics, King's College London, Strand,
London WC2R 2LS, UK}
\address{$\ddag$ ~ Department of Informatics, King's College London, Strand,
London WC2R 2LS, UK}
\address{$\S$ ~ Theoretical Physics, School of Physics and Astronomy, The University of Manchester, Manchester, M13 9PL, UK}

\ead{aleksandra.aloric@gmail.com}

\begin{abstract}
Loyal buyer-seller relationships can arise by design, e.g. when a seller tailors a product to a specific market niche to accomplish the best possible returns, and buyers respond to the dedicated efforts the seller makes to meet their needs. We ask whether it is possible, instead, for loyalty to arise spontaneously, and in particular as a consequence of repeated interaction and co-adaptation among the agents in a market. We devise a stylized model of double auction markets and adaptive traders that incorporates these features. Traders choose where to trade (which market) and how to trade (to buy or to sell) based on their previous experience. We find that when the typical scale of market returns (or, at fixed scale of returns, the intensity of choice) become higher than some threshold, the preferred state of the system is segregated: both buyers and sellers are segmented into subgroups that are persistently loyal to one market over another. We characterize the segregated state analytically in the limit of large markets: it is stabilized by some agents acting cooperatively to enable trade, and provides higher rewards than its unsegregated counterpart both for individual traders and the population as a whole.
\end{abstract}

\section*{Introduction}

When customers develop a persistent loyalty to one group of competing suppliers, or when suppliers target a specific segment of their potential customer base, we regard this as an instance of segregation of traders. In each case the interaction is between fixed (or largely fixed) groups of buyers and sellers and each group is segregated into subgroups: buyers have specific needs and these are met by specific products or suppliers. Such segregation can confer benefits, e.g. shorter exploration time when buying, but it may also add vulnerability to a system, for example by reducing market liquidity.

Segregation-like phenomena have been observed in a number of studies; examples we are aware of include emergent loyalty in a fish market~\cite{fishmarket}, segmentation in credit markets in the Philippines~\cite{segmentationPhilippines}, persistence in customer visitation patterns~\cite{Krumme2013} and specialization and herding in financial markets~\cite{specherding}. Some of the previous work has focused on the buyer side, and has studied how segmentation can be identified in the space of customers, and how suppliers may respond to it adequately once it is detected~\cite{customerclustering, CATcosegregation}. This branch of work can be thought of as a continuation of Adam Smith's idea that specialization of production leads to better exploitation of the market~\cite{ASmith}. Apparent paradoxes have also emerged, such as \textit{Hotelling's law}~\cite{Hotelling}: contrary to popular belief rational trading strategies may lead to minimal differentiation. In Hotelling's original example sellers that minimise their total distance from all potential customers, who are uniformly positioned along the beach, end up positioning themselves all in the same place -- in the middle of the beach.

Much less work has been done to explore the conditions under which segregation occurs in markets, and what factors promote or suppress it. When the buyer and seller populations are heterogeneous due to external factors such as geographical location or individual wealth, this can be a natural trigger for segregation.
Here, our aim is to investigate whether and when segregation can emerge {\em spontaneously} in initially homogeneous populations. In an \textit{in silico} experiment based on CAT Market Design tournaments~\cite{CAToverview} such segregated states were observed in the co-adaptation of markets and traders. Some markets ended up focusing on attracting specific traders, and some traders preferred to trade at a specific market although no such fixed propensities were imposed from the beginning.

Outside the context of double auction markets, an effect similar to segregation was noticed in multi-asset minority games~\cite{Huang2012}. Specifically, in a system of agents who have the option of copying a winning (minority) strategy from their neighbourhood with some probability $p$, a self-organized grouping of strategies arises when the probability $p$ is large enough. Grouping was only found when the underlying interaction network is well connected, and when agents have perfect information about the success of their neighbours. In another study based on the minority game~\cite{segregation99} it was noted that ``\ldots a population of competing agents with similar capabilities and knowledge will tend to self-segregate into opposing groups characterized by extreme behaviour''. The variant of the minority game used in this study supposes that agents can choose to follow a  global strategy based on collective memory -- this is the primary driver of segregation. Our approach is simpler in that the agents learn entirely from their own market returns and the only interaction between agents is via the market.

While the CAT tournament results referred to above provide a clear impetus to study spontaneous segregation, they are difficult to analyse theoretically due to the complexity of both the agents' trading strategies and the market mechanisms of the competing markets. We therefore devise and analyse a stylized individual-based model of a double auction. We focus on agents using experience-weighted attraction learning (EWA)~\cite{EWA} to learn from the payoffs received for past actions and thus optimise future trading. This general framework is similar to the one employed by Hanaki et al~\cite{luckystar}, though in a rather different context. Hanaki et al show that when agents are repeatedly given a choice of parking spots, some \textit{learn to be lucky} by focussing their search on to the spots very close to the city center, while the others end up with the opposite strategy.

By developing a stylized model we aim to obtain an intuitive and analytically tractable tool for understanding whether and when segregation can emerge spontaneously in a system of competing double auction markets. In the Results section we first describe the Model and then provide Numerical Results, summarizing  the main findings from our numerical simulations. In the following subsection we develop an Analytical Description in the large market limit, using a Fokker-Planck approach to investigate the steady states of the system mathematically  and in particular to develop insights into the properties of the segregated state. In the Discussion section we summarize our main findings, discuss their robustness to e.g.\ variations in the model and set out possible directions for future research. To the best of our knowledge, there is no other research that models spontaneous segregation in the context of double auction markets.

\section*{Results}

\subsection*{Model of Learning at Double-Auction Markets}

To address the question of spontaneous segregation, we study a stylized model of a population of adaptive traders and two double auction markets. We hypothesize that segregation can arise as a product of co-adaptation of traders, and construct on this basis a minimal model for both traders and markets. We investigate this in detail by numerical simulation, and provide a full theoretical understanding and characterisation of the observed segregation effects.
A brief summary of the basic features of the model and initial simulation results can be found in ~\cite{AEpublication}.

\paragraph{Traders.} Following the works of Gode and Sunder~\cite{ZItraders} and Ladlay~\cite{ZIladley}, we populate our system with agents without sophisticated trading strategies, essentially zero-intelligence traders. This is done because our aim is to investigate whether market loyalty can arise as an intrinsic property of a system of interacting agents without reliance on complex trading strategies.  Trading operates in discrete rounds, where at each turn the prices of buy and sell orders (bids and asks) that the traders submit to a market are independent identically distributed random variables (thus Zero-Intelligence agents). Agents learn whether to buy or to sell from previous experience (see below). The numerical values of bids and asks (i.e their magnitudes), however, are assumed to be unrelated to previous trading success or any other information. We assume that bids, $b$, and asks, $a$, are normally distributed ($a \sim \mathcal{N}(\mu_a,\sigma_a^2)$ and $b \sim \mathcal{N}(\mu_b,\sigma_b^2)$), and that their means satisfy $\mu_b>\mu_a$. In the spirit of the work of Gode and Sunder ~\cite{ZItraders}, $a, b$ can be thought of as cost and redemption values for each trader. These values are private to each trader but correspond to order prices if submitted bids and asks are truthful and reflect the actual valuation of goods by the agents. Traders will not accept any price from the market: sellers will trade only if the trading price $\pi$ is no less than their asking price ($\pi \geq a$), and buyers require that the trading price is no greater than what they bid ($\pi \leq b$). (This is in line with Gode and Sunder~\cite{ZItraders} where the traders that turned out to be more similar to human traders were the zero-intelligence ones with budget constraint, i.e.\ traders who were not allowed to trade at loss -- higher than the redemption value for buyers and lower than the cost value for sellers.) After each round of trading each agent receives a score, reflecting their payoff in the trade. The scores of agents who do trade are assigned as follows: buyers value paying less than they offered ($b$), and so their score is $S=b-\pi$. Sellers value trading for more than their ask ($a$), and so  $S=\pi-a$ is a reasonable model for their payoff. We note that these scores are based on a linear model and do not reflect effects such as diminishing returns. Traders who do not get to trade in a given round receive return $S=0$.

\paragraph{Markets.} The role of a market is to facilitate trades so we define markets in terms of their price-setting and order matching mechanisms; for an in-depth review of possible double auction market mechanisms see~\cite{DAexplained}. We consider a single-unit discrete time double auction market where all orders arrive simultaneously and market clearing happens once every period after the orders are collected. (Note that each period consist of one round only, which is why we will talk only in terms of periods.)

\begin{figure}[h!]
\vspace{-10pt}
    \includegraphics[width=\textwidth]{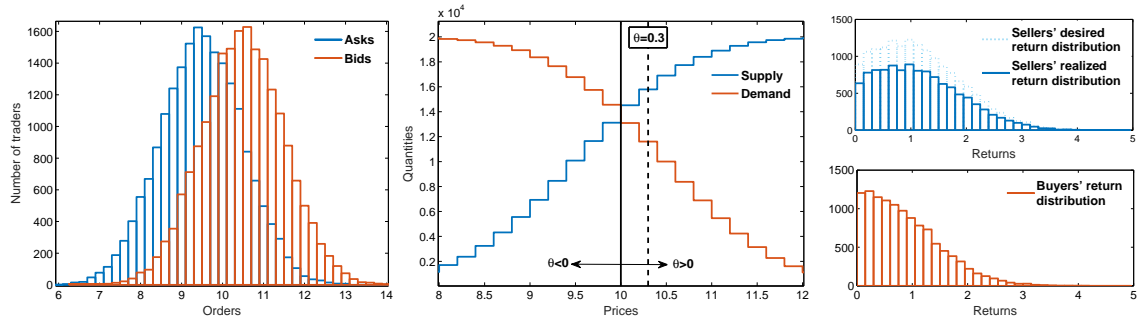}
    \caption{{\bf Illustration of the market mechanism.} Left: example distributions of orders received in one trading period (orange -- orders to buy, blue -- orders to sell). Middle: the corresponding demand and supply curves, showing the number of orders to buy above (resp.\ sell below) a certain price. The vertical black line marks the equilibrium trading price, i.e.\ the price at which demand equals supply, while the arrows show how the trading price changes for nonzero $\theta$; the dashed line is the trading price for $\theta=0.3$. Right: example distributions of returns from validated orders at $\theta=0.3$. In the situation shown, where buyers are in the minority after validation of orders, all buyers find a trading partner (red). If all sellers could trade they would achieve the ``desired'' return distribution (light blue). But only a randomly selected fraction finds a trading partner, leading to the ``realized'' return distribution (dark blue); the other sellers receive a return of zero. The realized sellers' return distribution has the same total area as the buyers' return distribution as both represent the total number of trades.}
    \label{fig1:marketmech}
\end{figure}

We also assume that a uniform price is set by the market -- once all orders have arrived, these are used to determine demand and supply (see Fig.~\ref{fig1:marketmech}). The price at which demand equals supply is the equilibrium price. To make the model more flexible, we consider more generally uniform prices of the type
\begin{equation} 
\pi=\pi^{\rm eq}+\theta(\langle b\rangle-\langle a\rangle)
\label{price}
\end{equation} where $\pi^{\rm eq}$ is the equilibrium trading price, from which the market price can deviate towards the average bid ($\langle b\rangle$) or the average ask ($\langle a\rangle$); the parameter $\theta$ thus represents the bias of the market towards buyers or sellers. In a setup like ours where the bids and asks are Gaussian random variables, the equilibrium trading price is $\pi^{\rm eq}=(\langle b\rangle+\langle a\rangle)/2$  when the variances $\sigma_a$ and $\sigma_b$ are equal as we assume here and below. Once the trading price has been set, all bids below this price, and all asks above it, are marked as invalid orders as they cannot be executed at the current trading price. The remaining orders are executed by randomly pairing buyers and sellers; the execution price is $\pi$. Note that we assume here that each order is for a single unit of the good traded.

An example of a discrete time double auction market is the Opening Auction of New York Stock Exchange which is used to determine the opening prices on the market. At the opening auction, once all the orders are submitted, the trade occurs at the single price set by the market that maximises the volume of trades. In the terminology of our model, this price corresponds to $\theta=0$ as that setting is the one that maximises the number of possible trades.

Fig.~\ref{fig1:marketmech} shows graphically how the trading price is set and how the return distributions look after invalid bids and asks are eliminated. If, as in the example in the figure, the trading price is closer to the average bid ($\theta>0$) there will be an excess of supply: only a fraction of the sellers who submitted valid asks gets to trade, but with the higher trading price -- so those that do trade receive higher returns. The buyers who submitted valid bids are in a minority compared to the sellers and so determine the number of trades. They can all trade, but receive relatively low returns on account of the high trading price.

\paragraph{Reinforcement Learning Rule.} So far we have described how agents interact at a given market. We next define how they decide \textit{how to trade} (to buy or to sell), and \textit{where to trade} (at which market). Agents trade repeatedly in our model, and they adapt their preferences for the various choices from one trading period to the next. We focus on the case of two markets, though the model can easily be extended to an arbitrary number of markets.

We assume that each agent decides on an action at the beginning of each trading period, only based on his or her past experience. To formalize this we introduce  a set of attractions $A_\gamma$ for each player, one for each action $\gamma$: buy at market 1, sell at market 1, buy at market 2, sell at market 2. The attractions will generally differ from player to player, but we suppress this in the notation for now. The attractions are updated after every trading period, $n$, using the following reinforcement rule:
\begin{equation}
A_{\gamma}(n+1) = \begin{cases}
(1-r)A_{\gamma}(n) + rS_{\gamma}(n), & \mbox{if the agent chose action $\gamma$ in round $n$}\\
(1-r)A_{\gamma}(n), & \mbox{if the agent chose an action $\delta\neq\gamma$ in round $n$}.
\end{cases}
\label{reinfrule}
\end{equation}
The quantity $S_{\gamma}(n)$ is the score gained by taking action $\gamma$ in the $n$-th trading period. The length of the agents' memory is set by $r$: agents weight returns obtained $\Delta n$ time steps ago with exponentially decaying factors $(1-r)^{\Delta n}$, effectively corresponding to a sliding window of length of order $1/r$ for the weighted averaging of past returns.

The update rule above is a special instance of a more general experience-weighted attraction rule ~\cite{EWA,EWAnew}, which has been shown to be in reasonable agreement with experimental data on human learning in repeated games. Many special cases of this rule are in common use in evolutionary biology and in the game-theoretic literature. 
One important variant of EWA is a case in which {\em all} actions are updated with their returns, no matter whether that action was actually taken or not. This assumes that an agent can calculate or at least estimate the return (s)he would have obtained from actions $\gamma$ that (s)he did not choose to play~\cite{CrutchfieldSato2003, GallaFarmer2013}. We argue that an agent would not normally have sufficient information to do this in the context of a double auction market: (s)he would need access to the current price, and to the numbers of valid bids and asks submitted. This is unrealistic, which is why we posit that scores of unplayed actions are updated with an effective payoff of zero. One plausible alternative that does {\em not} rely on estimation of returns from unplayed actions would be an update rule where the attractions to unplayed actions are forgotten with a different rate (or not forgotten at all); we return to this in the Discussion at the end.

It only remains to specify how agents choose their actions based on the attractions. This is done in line with the experience-weighted attraction literature \cite{EWA,EWAnew}, simply by converting the attractions to probabilities using the so-called \textit{softmax} or \textit{logit} function. Explicitly, each agent takes action $\gamma$ in trading period $n$ with probability $P({\gamma}|\mathbf{A}(n)) = \exp{(\beta A_{\gamma}(n))}/\sum_{\gamma'}\exp{(\beta A_{\gamma'}(n))} \propto \exp{(\beta A_{\gamma}(n))}$, where $\beta$ is the \textit{intensity of choice}~\cite{GallaFarmer2013} and regulates how strongly the agents bias their preferences towards actions with high attractions. For $\beta\rightarrow\infty$ the agents choose the option with the highest attraction, while for $\beta\rightarrow 0$ they choose randomly with equal probabilities among all options. One way to interpret $\beta$ is as parameterizing the degree of human rationality, where $\beta\rightarrow\infty$ corresponds to the limit of unboundedly rational players who always choose the optimal course of action~\cite{Goeree}. Another interpretation is that $1/\beta$ sets a scale of return differences below which traders no longer significantly differentiate between the options available to them. Indeed, if all $A_\gamma(n)$ are within $1/\beta$ of each other then the exponents in the softmax function differ by less than unity and so the resulting probabilities are close to uniform. The existence of such a threshold is not implausible: higher return differences should drive a trader towards optimizing his or her actions more, e.g.\ by following the historically most rewarding options, while the choice between actions giving nearly the same return will be largely random. The intensity of choice $\beta$ is sometimes denoted $\lambda$ and has also been referred to as \textit{response sensitivity}~\cite{EWA,EWAnew,luckystar} or \textit{learning sensitivity}~\cite{CrutchfieldSato2003}.

\subsection*{Numerical Results}

{\bf Emergence of loyalty - return-oriented and volume-oriented traders.} We begin our exploration of the two-market setup defined above with results from numerical simulations. Unless specified otherwise, parameters were set as detailed in Table \ref{parametertable} in the Methods section. All agents start with equal preferences for all four choices, i.e.\ $A_\gamma=0~ \forall \gamma \in\{\mathcal{B}1,\mathcal{S}1,\mathcal{B}2,\mathcal{S}2\}$ where $\mathcal{B}=$ buy and $\mathcal{S}=$ sell. To aid visualization we project the four-dimensional space of attractions $A_\gamma$ down to two coordinates, the overall attraction to buying as against selling, defined as $\Delta_{\mathcal{BS}}=(A_{\mathcal{B}1}+A_{\mathcal{B}2})-(A_{\mathcal{S}1}+A_{\mathcal{S}2})$, and the attraction to market 1 as against market 2, $\Delta_{12}=(A_{\mathcal{B}1}+A_{\mathcal{A}1})-(A_{\mathcal{B}2}+A_{\mathcal{S}2})$. Due to the nonlinearity of the \textit{softmax} function, a single-peaked distribution of attractions with a non-zero spread will, for large enough intensity of choice $\beta$, become a multimodal distribution in the space of preferences. This effect cannot be regarded as genuine segregation. For this reason we will avoid representing agents in the space of their preferences $P_\gamma$, and use the underlying attractions instead.

\begin{figure}[h!]
 \vspace{-5pt}
        \includegraphics[width=\textwidth]{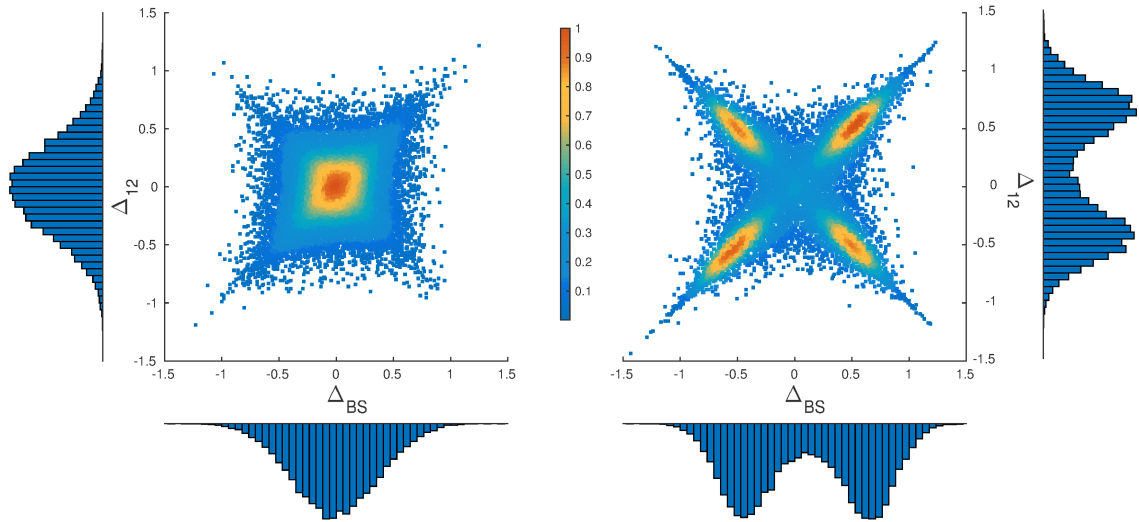}
        \caption{{\bf Steady states in numerical simulations.} Steady state distributions of attractions for buying vs selling and for market 1 vs 2, respectively, for forgetting rate $r=0.1$. The colour scale in the central scatter plots shows the probability density of agents\cite{scatter}. For low enough intensity of choice (left, $\beta=3.45$), an unsegregated steady state arises, while higher $\beta$ (right, $\beta=7.14$) causes traders to segregate into four discernible groups. The histograms along the axes show the marginal distributions of $\Delta_{\mathcal{BS}}$ and $\Delta_{12}$, respectively. From the marginals we note that in the segregated state roughly half of the population specialize in buying (similarly half prefers market 1), although the fractions of buyers (sellers) at different markets are not the same. Other system parameters are given in Table \ref{parametertable}. The data shown are taken from 100 trading periods of 100 independent runs.}
\label{fig2:simulation}
\end{figure}

In Fig.~\ref{fig2:simulation} (left) we present the steady state attraction distribution for a population of traders with intensity of choice $\beta=3.45$. The initially narrow, \textit{delta-peaked} distribution of attractions (all initialized at $0$) has been broadened due to diffusion arising from the random nature of returns and from the stochasticity of the agents' actions. The steady state shown in the left panel of Fig.~\ref{fig2:simulation} represents an unsegregated population of traders. While this population does include some traders with moderately strong preferences for one of the actions, preferences remain weak on average. The population as a whole remains homogeneous in the sense that there is no split into discernible groups.

Fig.~\ref{fig2:simulation} (right) contrasts this scenario with the steady state of a system with exactly the same set of parameters but at the higher intensity of choice $\beta=7.14$. The population of traders now splits into four groups, with the agents \textit{persistently trading} preferentially at one of the markets and with a preferred buy or sell action. We refer to this state as \textit{segregated}. The markets shown in this example (Fig.~\ref{fig2:simulation}) are biased symmetrically ($\theta_{1,2}=\mp 0.2$):  if an agent buys at market 1, or sells at market 2 (actions $\mathcal{B}1$ or $\mathcal{S}2$), and if they manage to trade, they are awarded with a higher score on average than if they were to choose one of the other two actions (cf.\ Fig.~\ref{fig1:marketmech}, where sellers are the group with the higher average return). In this symmetric market setup there are therefore two different kinds of behaviour among the four segregated groups. The agents in two of the groups ($\mathcal{B}1$ and $\mathcal{S}2$) specialize in actions that award them higher average return, we call those traders \textit{``return-oriented''}. We will refer to the agents in the other two groups ($\mathcal{B}2$ or $\mathcal{S}1$) as\textit{``volume-oriented''} for reasons which we explain shortly. If all traders were return-oriented, they would have no partners to trade as there would be no agents playing $\mathcal{B}2$ or $\mathcal{S}1$. Consequently everyone would receive zero returns. The fact that some traders have developed persistent preferences for placing orders that give them a lower average return (here $\mathcal{B}2$, $\mathcal{S}1$) can thus be interpreted as cooperative, trade-enabling behaviour. Due to the market mechanism, the agents who choose one of $\mathcal{B}2$ or $\mathcal{S}1$ will have their orders rejected as invalid more often (see the group of buyers in Fig.~\ref{fig1:marketmech}). However, those whose orders are accepted will form a minority group in the market and will always find a trading partner. Hence, when submitting valid orders, these traders trade more often so we will call them \textit{``volume-oriented''}.

There is a tendency for agents to cluster around the diagonal and anti-diagonal in the two-dimensional projection shown in Fig.~\ref{fig2:simulation}. Inspection of the underlying attractions $A_\gamma$ shows that this arises because many agents have one large attraction, for their preferred action, while their other attractions are close to zero. For example if a return-oriented agent, say a buyer at market 1, has attractions $\mathbf{A}\approx(A_{\mathcal{B}1},0,0,0)$, then this projects to the diagonal $\Delta_{\mathcal{BS}}=\Delta_{12}=A_{\mathcal{B}1}$. Similarly a volume-oriented trader who prefers to sell at market 1 is projected to the anti-diagonal $-\Delta_{\mathcal{BS}}=\Delta_{12}=A_{\mathcal{S}1}$ if his/her other three attractions are close to zero.
The fact that the attractions of non-preferred actions are often small comes from the fact that, within our reinforcement learning dynamics, agents gradually forget the scores of actions they only use rarely.
\\

{\bf Persistence and time correlation.}
The above results indicate that segregation is seen above a critical intensity of choice, $\beta>\beta_s$; we defer a discussion of how $\beta_s$ depends on the parameters of the model to the section Analytical Description below. One has to bear in mind though that the data shown in Fig.~\ref{fig2:simulation} represents the state of the system at a given time. It therefore does not tell us whether agents really develop loyalty in the sense that they stay in one of the segregated groups for a long time, or whether they switch frequently between groups.

To understand which of these two alternatives applies, and to characterise the dynamics of group switching events, one can analyse the persistence times of agents within each of the four peaks in Fig.~\ref{fig2:simulation}. Consistent with the intuitive meaning of a segregated state one finds (as shown in \cite{AEpublication}) that at high enough intensity of choice $\beta$ the agents develop ``loyalty'' to a certain market and a choice of buying and selling: their persistence times are much longer than the timescale of the small short-term fluctuations in preferences that every agent experiences. These short-term fluctuations occur on the time scale of the memory-loss, $1/r$. On the other hand the agents are {\em not} frozen, i.e.\ the persistence times are finite and the agents change loyalties on longer timescales. Therefore the steady state we observe is well defined rather than a consequence of agent preferences frozen-in from early fluctuations in their trading history.

\begin{figure}[h!]
       \centering
             \includegraphics[width=\textwidth]{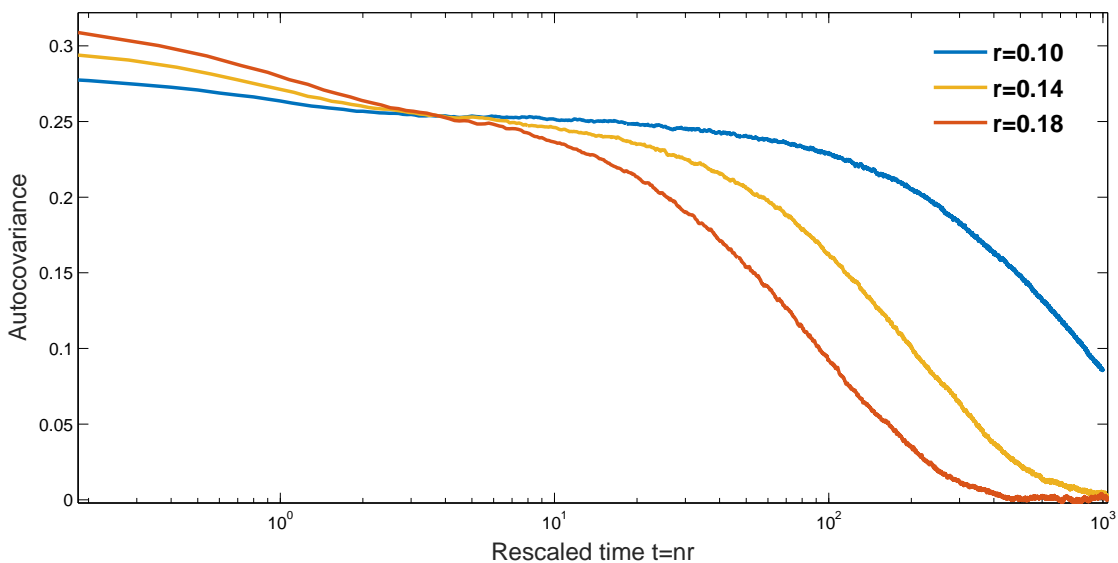}
                \caption{{\bf Decay of the attraction autocovariance with time.} The intensity of choice $\beta$ and forgetting rate $r$ are different from the standard values in Table \ref{parametertable}: here $\beta=20$, $r$ as shown in the legend. This is to highlight the separation of timescales in the segregated state. A larger than usual initial number of trading periods was simulated ($\approx20,000$ trading periods) to ensure the systems reached steady state at such high $\beta$.}
        \label{fig3:covariance}
\end{figure}

What emerges from the above discussion is that a key feature of a segregated state is a separation of timescales in the dynamics of the agents. To quantify this further we consider the autocovariance function
\begin{equation}
C_{\gamma \delta}(\tau)=\langle (A_\gamma(t_0+\tau)-A_\gamma(t_0))(A_\delta(t_0+\tau)-A_\delta(t_0))\rangle_{P(\mathbf{A})},
\end{equation}
where $\langle \dots \rangle_{P(\mathbf{A})}$ indicates an average over the attractions of players in the population, and over the time $t_0$ in the stationary state.
This is a matrix capturing cross-correlations in time between the attractions to the various actions an agent can take; here as a summary statistic we look at the trace $C(\tau)\equiv\sum_\gamma C_{\gamma\gamma}(\tau)$, i.e.\ the sum of the autocovariances. In Fig.~\ref{fig3:covariance} we show how this autocorrelation function depends on time for various forgetting rates $r$ at fixed $\beta=20$. We observe two separate decays, which is consistent with the intuition described above: first a fast decay of correlations occurs as an agent moves around one of the peaks of the distribution $P({\bf A})$, caused by randomness in returns at a given market and the fact that the decisions of each agent remain stochastic. This fast decay does not lead to full decorrelation but rather to a plateau in the autocorrelation function: the attractions of an agent stay in the same ``loyalty group" (i.e. within the same peak of $P({\bf A})$), and the agent's actions remain correlated. The decay from this plateau defines the slow timescale, and it is a measure of how long agents typically stay in one loyalty group before moving to another peak of $P({\bf A})$.

Because the slow timescale increases very rapidly as $r$ is decreased -- the theoretical description below suggests exponential growth with $1/r$ -- we show a rather smaller range of $r$ in the figure than elsewhere in the paper. Comparing the curves for different $r$ we note the increase of the initial value of the autocorrelation function with $r$, which makes sense as the variance of the fluctuations within each peak of $P({\bf A})$ grows roughly proportional to $r$. We also see that the fast time scale is of the order of unity in the rescaled time units, $t=nr$, used in the figure. This reflects the fact that the effective memory of each agent is $1/r$ trading periods.

{\bf Summary and characterisation of segregated state.} Summarizing so far, the existence of a plateau in the autocorrelation function is clear evidence of the segregation of agents into groups that remain loyal over extended periods of time to a certain set of preferences defining a group of agents, while at the same time exhibiting small and much faster fluctuations around these typical preferences. This represents the most intuitive definition of segregation that we can think of. However, the existence of a plateau -- quantitatively, a turning point in a plot of autocorrelation versus log time -- would not be easy to utilize in practice to detect segretation. This is because already for $r<0.1$ the amplitude of the initial fast decay becomes so small that it merges into the plateau, while at the same time the slow timescale for switching between loyalty groups outgrows the range of times that can easily be explored computationally. For practical purposes we therefore stick to multimodality of $P({\bf A})$ as our criterion for segregation.

Our simulation results suggest that even our simplified trading system shows rich and interesting behaviour. A more detailed analysis (discussed below) reveals a threshold intensity of choice $\beta_{\rm s}$ such that for values $\beta>\beta_{\rm s}$ the system segregates, i.e.\ the initially homogeneous population of traders splits into groups that persistently choose to trade at a specific market. The pattern of these market loyalties is co-operative, in that some traders -- the volume-oriented ones -- forego potentially higher returns and instead choose actions that enable trade for everyone. We will see below that this behaviour has clear payback, in giving average returns across all traders that exceed those for the obvious Nash equilibrium. An obvious question that remains open at this point is how robust the observed segregation behaviour is to variation in our model parameters and setup; we defer this issue to the discussion section.

\subsection*{Analytical Description}

\paragraph{Fokker-Planck description.} To understand in more detail how segregation arises, and to characterise the nature of the transition to the segregated state, a theoretical analysis of the model dynamics would evidently be useful. Mathematically our model is Markovian. To capture the full dynamics one needs to keep track of the attractions $A_\gamma^i$ to all actions $\gamma\in\{\mathcal{B}1, \mathcal{S}1, \mathcal{B}2, \mathcal{S}2\}$ of all agents $i=1,\ldots,N$, or equivalently the attraction vectors $\mathbf{A}^i=(A^i_{\mathcal{B}1},A^i_{\mathcal{S}1},A^i_{\mathcal{B}2},A^i_{\mathcal{S}2})$ for all agents. One can write down a master equation for the evolution of the joint distribution of these $4N$ variables. This equation is however difficult to work with and not easily analysed further. In order to make progress it is useful to realise that the payoff an agent receives when they take a particular action only depends on aggregated quantities, but not on the detailed actions of individual other agents. More specifically the return for a given action is determined by (i) the bid or ask the agent places; (ii) the validity of the bid or ask placed and (iii) whether or not a suitable trading partner is found. All quantities (i)-(iii) are random objects and so the return received for a given action at any one time will be a random variable itself. Properties (ii) and (iii) only depend on the macroscopic statistics of bids and asks placed by the population of traders in its entirety. We now focus on the limit of large (formally infinite) populations, that is we take the limit $N\to\infty$. The distribution of bids and asks submitted in any trading round will then follow Gaussian distributions with mean $\mu_a$ and standard deviation $\sigma_a$ for asks, and analogously for bids (${\cal N}(\mu_b,\sigma_b)$). By definition in our model the returns that flow from these bids or asks are non-negative, implying a truncation at zero of the return distributions. These distributions can then be written as

\begin{align}
P(S|m,\mathcal{B})
& =
Q_{\mathcal{B}m}T_{\mathcal{B}m}
\frac{1}{Q_{\mathcal{B}m}\sigma_b\sqrt{2\pi}}
\exp\left(-\frac{(S-\pi_m)^2}{2\sigma_b^2}\right)
\theta(S)
+\delta(S)(1-Q_{\mathcal{B}m}T_{\mathcal{B}m})\nonumber\\
P(S|m,\mathcal{S})& =
\underbrace{Q_{\mathcal{S}m}T_{\mathcal{S}m}}_\text{agent trades}\underbrace{\frac{1}{Q_{\mathcal{S}m}\sigma_a\sqrt{2\pi}}
\exp\left(-\frac{(S-\pi_m)^2}{2\sigma_a^2}\right)
\theta(S)}_\text{non-negative return}
+\delta(S)\underbrace{(1-Q_{\mathcal{S}m}T_{\mathcal{S}m})}_\text{agent does not trade}.
\label{scores}
\end{align}
The return an agent receives depends on the action $\gamma$ the agent chooses, i.e. on  the market, $m$, they trade in, and whether they chose to buy, $\mathcal{B}$, or sell, $\mathcal{S}$. This is reflected in our notation above.

The first term in each of the above expressions describes the case of a non-zero return. This occurs with probability $Q_\gamma T_\gamma$, where
 $Q_\gamma$ denotes the probability that an agent's order is valid; once validated, it is executed with probability $T_\gamma$, which is the probability of finding a suitable trading partner. If the order is executed an agent receives a return $S$ drawn from a Gaussian distribution, truncated to allow only non-negative payoffs; if the order was invalid, or was valid but not executed, the agent receives $S=0$. This occurs with probability $1-Q_\gamma T_\gamma$.

To complete the above description of the single agent dynamics for large $N$ one needs expressions for the market price $\pi_m$ as well as the probabilities $Q_\gamma$ and $T_\gamma$. In the deterministic limit, $N\to\infty$, the expression in Eq. (\ref{price}) reduces to
\begin{equation*}
\pi_m=\pi^{\rm eq}+\theta_m(\mu_b-\mu_a)\qquad\text{with }\pi^{\rm eq}=\frac{\mu_a+\mu_b}{2},
\end{equation*}
as the means of the bids and asks submitted in any one round become the population means $\mu_b$ and $\mu_a$ by virtue of the law of large numbers. In the case when $\sigma_a=\sigma_b$, the bid and ask distributions are mirror images of each other around $(\mu_a+\mu_b)/2$ and so the quantity $\pi^{\rm eq}$ introduce above is the equilibrium price where demand equals supply. (For $\sigma_a\neq \sigma_b$, and in general when the number of buyers is not equal to the number of sellers at a given market, the real equilibrium trading price is not given by this simplified formula, but in our model the market price is still calculated from average bids and asks and we retain the symbol $\pi^{\rm eq}$ for the average of the two means.)

The probabilities that an order is valid, $Q_\gamma$, are calculated from
\begin{align}
Q_{\mathcal{B}m}&=\frac{1}{\sigma_b\sqrt{2\pi}}\int_{\pi_m}^{\infty}db \,\exp\left(-\frac{(b-\mu_b)^2}{2\sigma_b^2}\right)\nonumber\\
Q_{\mathcal{S}m}&=\frac{1}{\sigma_a\sqrt{2\pi}}\int_{-\infty}^{\pi_m}da \,\exp\left(-\frac{(a-\mu_a)^2}{2\sigma_a^2}\right).
\label{validorder}
\end{align}
These expressions reflect the requirement for a bid or ask to be on the correct side of the market price, and are based on our assumption of Gaussian bid and ask distributions. The integrals can be carried out in closed form and expressed in terms of error functions.

The trading probabilities $T_{\gamma}$, finally, can be written as
\begin{equation}
  \begin{split}
    T_{\mathcal{B}m}=\frac{\min\left(\overline{N}_{\mathcal{B}m},\overline{N}_{\mathcal{S}m}\right)}{\overline{N}_{\mathcal{B}m}}
  \end{split}
  \qquad\qquad
  \begin{split}
 	T_{\mathcal{S}m}=\frac{\min\left(\overline{N}_{\mathcal{B}m},\overline{N}_{\mathcal{S}m}\right)}{\overline{N}_{\mathcal{S}m}}
  \end{split}
  \label{tradingP}
\end{equation}
where $\overline{N}_\gamma$ is the total number of agents taking an action $\gamma$ and submitting a valid order, 
\begin{equation}
\overline{N}_{\gamma}=NQ_\gamma\left\langle P(\gamma|\mathbf{A})\right\rangle.
\label{NumberOfTraders}
\end{equation}
In this expression $P(\gamma|\mathbf{A}) \propto \exp(\beta A_\gamma)$ is given by the appropriate \textit{softmax} function applied to an agent's vector of attractions $\mathbf{A}$. The average $\langle\dots\rangle$ is over the distribution $P(\mathbf{A})=\frac{1}{N}\sum_{i=1}^N \delta(\mathbf{A}-\mathbf{A^i})$ of attraction vectors across all agents.
The expressions (\ref{tradingP}) can be understood as follows: if a trader submits an order that is valid, (s)he will always be able to trade if (s)he is in the minority group, otherwise his/her probability of being able to trade is the ratio of the number of traders in the minority and majority groups.

We can now write an evolution equation for the distribution $P(\mathbf{A})$ of attraction vectors across the population of traders.
Writing $P_n(\mathbf{A})$ for the distribution at the end of trading period $n$, we have
\begin{equation}
 P_{n+1}(\mathbf{A}')=\int d\mathbf{A}\, K(\mathbf{A}'|\mathbf{A})P_n(\mathbf{A})
\label{masterEq}
 \end{equation}
with $K(\mathbf{A'}|\mathbf{A})$ a transition kernel that encodes the dynamics of the system. It is of the form
 \begin{equation}
 K(\mathbf{A}'|\mathbf{A})= \int dS
 \,\sum_{\gamma}
 P(S|\gamma)P(\gamma|\mathbf{A})\delta(\mathbf{A}'-\mathbf{e}_{\gamma}rS-(1-r)\mathbf{A}),
 \label{kernel}
 \end{equation}
where $\mathbf{e}_{\gamma}$ is a four-dimensional vector with an entry of $1$ for action $\gamma$ and entries $0$ otherwise. The learning rule of Eq. (\ref{reinfrule}) is enforced through the delta function. It is worth noting that (\ref{masterEq}) is not a standard linear Chapman-Kolmogorov equation as the right-hand side is nonlinear in the distribution $P_n(\mathbf{A})$. This arises because the kernel $K$ depends on the trading probabilities $T_\gamma$ as given in (\ref{tradingP}), which in turn depend on $P_n(\mathbf{A})$. The nonlinearity arises because we have effectively projected from a description in terms of all $4N$ attractions to one involving only four single-agent attractions.

From our reasoning so far, Eq. (\ref{masterEq}) should constitute an exact description of the model in the limit $N\to\infty$. It can, at least in principle, be solved numerically starting from our chosen initial condition $P_0(\mathbf{A})=\delta(\mathbf{A})$. The presence of the $\delta$-peaks at zero returns $S=0$ makes the kernel awkward to deal with numerically, however. We therefore make one further simplification and transform to a Fokker-Planck description. This is appropriate for small $r$, i.e.\ agents with long memory. Note to this end that the change in attraction $\mathbf{A}'-\mathbf{A}$ in any one trading period is directly proportional to $r$, see Eq. (\ref{kernel}). The Kramers-Moyal expansion (see e.g. \cite{Kampen}) of equation (\ref{masterEq}) is of the form
\begin{equation}
 P_{n+1}(\mathbf{A})-P_n(\mathbf{A})
  =
- r \partial_\mathbf{A} \left[\mathbf{M}_1(\mathbf{A})P(\mathbf{A})\right]
+\frac{r^2}{2}\partial_\mathbf{A}^2 \left[\mathbf{M}_2(\mathbf{A})P(\mathbf{A})\right] + \ldots,
\label{KM}
\end{equation}
where the scaled jump moments
\begin{equation*}
\mathbf{M}_\ell(\mathbf{A})=\frac{1}{r^\ell}\int d\mathbf{A'} (\mathbf{A'}-\mathbf{A})^\ell K(\mathbf{A'}|\mathbf{A})
\end{equation*}
are of order $r^0$. The generic $\ell$-th order term on the RHS of equation (\ref{KM}) comes with a factor $r^\ell$ so for small $r$ one can proceed by neglecting higher-order terms beyond the first two. Next, it is useful to introduce a re-scaled time $t=rn$. A unit time interval in $t$ then corresponds to $1/r$ trading periods and hence the memory length of the agents. With this replacement, Equation (\ref{KM}) reduces to a Fokker-Planck equation in the limit $r\to 0$. Specifically we have in this limit
\begin{equation}
\partial_t P(\mathbf{A})=
-\partial_\mathbf{A}\left[\mathbf{M}_1(\mathbf{A})P(\mathbf{A})\right]
+\frac{r}{2}\partial_\mathbf{A}^2\left[\mathbf{M}_2(\mathbf{A})P(\mathbf{A})\right].
\label{AdaptiveFokkerPlanck}
\end{equation}
As before and to keep the notation compact, we have not written out the various components of the derivatives and jump moments; e.g.\ $\partial_\mathbf{A}[\mathbf{M}_1 P]$ is to be read as $\sum_\gamma \partial_{A_\gamma}[M_{1,\gamma} P]$.

Segregation behaviour can in principle be characterised by studying the steady-state solution of the above Fokker-Planck Equation (\ref{AdaptiveFokkerPlanck}). Specifically one would investigate the conditions under which this distribution is multimodal, i.e.\ has several peaks. 
\paragraph{Iterative procedure for solving the Fokker-Planck equation.} Similar to the kernel $K$, the drift and diffusion coefficients $\mathbf{M}_1$ and $\mathbf{M}_2$  depend on the trading probabilities $T_\gamma$. So finding the steady state requires an iterative approach:\\
(i) initialize $P(\mathbf{A})$, e.g.\ with delta peaked distributions, corresponding to agents without preferences; \\(ii) calculate the number of traders taking the various actions [Eq.~(\ref{NumberOfTraders})] and thus the trading probabilities $T_\gamma$ [Eq.~(\ref{tradingP})];\\ (iii) find the steady state solution of the Fokker-Planck equation (\ref{AdaptiveFokkerPlanck}) for these $T_\gamma$. 
Steps (ii) and (iii) are then repeated until a self-consistent solution is obtained, i.e.\ until the $T_\gamma$ no longer change. We note that finding the stationary distribution of the Fokker-Planck equation (step (iii)) is non-trivial in general. This is true particularly because the drift and diffusion coefficients $\mathbf{M}_1$ and $\mathbf{M}_2$ do not define a time-reversible single agent-dynamics, for which determining the steady state would be much simpler (see for example~\cite{NonEq}). In the limit of small $r$, where the stationary distribution takes a large deviation form $P(\mathbf{A})\propto \exp(-f(\mathbf{A})/r)$ analogous to (\ref{stationarydistribution}) below, it can in principle be found by a Freidlin-Wentzel construction~\cite{FW} but implementing this numerically is still challenging (see e.g.~\cite{Nicole}). This is why for analytical work we focus on a slightly simplified model, details of which are given below.
Once we have the steady-state values of the trading probabilities $T_\gamma$, we can also think of the Fokker-Planck equation (\ref{AdaptiveFokkerPlanck}) as describing the dynamics of individual agents within a large population with fixed average properties. As we argue that our system is ergodic, the steady state distribution of attractions can then be re-interpreted as the distribution of a single agent's attraction sampled over a long enough time interval. Zeros of the drift velocity $\mathbf{M}_1(\mathbf{A})$ are fixed points of the single agent dynamics, and for small amounts of diffusive noise $r$ the single agent will spend most of its time near (stable) fixed points, causing local maxima of $P(\mathbf{A})$. To detect segregation we therefore look for multiple stable fixed points of the single agent-dynamics in the steady state population.

\paragraph{Agents with fixed preferences for buying.} As finding the steady state solution of the Fokker-Planck equation even with given trading probabilities is a non-trivial task in the four-dimensional space of attraction vectors $\mathbf{A}$, we proceed with one more simplification to produce a theoretical description directly comparable to simulations. We fix the agents' preferences for buying or selling, i.e. each agent now carries a fixed probability $p_{\mathcal B}$ with which they buy. They sell with probability $1-p_{\mathcal B}$.  This probability may vary from agent to agent, but crucially it remains fixed in time for each trader. Agents thus have a single decision left to make, namely, where to trade. In the case of two markets the single variable that we then need to track for every agent is the difference or relative attraction $\Delta^i=A^i_1-A^i_2$. This makes a full numerical analysis possible. The resulting Fokker-Planck equation is one-dimensional and so one can find the steady state solution in closed form, while locating all single agent fixed points can be achieved e.g.\ by a bisection method. We do find segregation in this way as shown below so the reduced model is useful in its own right, and provides evidence of the robustness of segregation behaviour. Hence we focus on this simplified model in the following, enabling us to compare numerical predictions for nonzero $r$ directly to simulations. Wherever possible we will relate the results back to the original, fully adaptive model.

In the limit of large population size, the analogue of $P(\mathbf{A})$ in the simplified model is the probability distribution $P(\Delta, p_\mathcal{B})=P(\Delta|p_\mathcal{B})P(p_\mathcal{B})$. The quantity $P(\Delta|p_\mathcal{B})$ is the distribution of attraction differences, $\Delta$, among agents with buying preference $p_{\mathcal B}$. The distribution over preferences for buying $P(p_\mathcal{B})$ is fixed as part of the specification of the model. As a simple case, we investigate a population with $P(p_\mathcal{B})=\frac{1}{2}\delta(p_\mathcal{B}-p_\mathcal{B}^{(1)})+\frac{1}{2}\delta(p_\mathcal{B}-p_\mathcal{B}^{(2)})$,
consisting of equal numbers of two types of agents with buying preference  
$p_{\mathcal B}=p_{\mathcal B}^{(1)}$ and $p_{\mathcal B}=p_{\mathcal B}^{(2)}$, respectively. Agent $i$ chooses market $1$ with probability $1/[1+\exp(-\beta\Delta_i)]$ and independently chooses to buy with probability $p_\mathcal{B}^{(1)}$ or $p_\mathcal{B}^{(2)}$ depending on his/her type.

As in the case of fully adaptive agents we can formulate the master equation for the process, and derive a Fokker-Planck equation in the limit of small memory-loss rates $r$. We obtain:
\begin{equation}
\partial_t P(\Delta|p_\mathcal{B}^{(g)})=
-\partial_\Delta\left[M_1(\Delta|p_\mathcal{B}^{(g)},T_\gamma)P(\Delta|p_\mathcal{B}^{(g)})\right]
+\frac{r}{2}\partial_\Delta^2\left[M_2(\Delta|p_\mathcal{B}^{(g)},T_\gamma)P(\Delta|p_\mathcal{B}^{(g)})\right]
\label{FokkerPlanck}
\end{equation}
where  $g\in \{1,2\}$ (for type or ``group'') labels the agent type. The jump moments $M_1$ and $M_2$ couple the two types of agents via the set of trading probabilities $\{T_\gamma\}$.
The generic steady state solution of the Fokker-Planck equation (\ref{FokkerPlanck}) reads (see for example \cite{Kampen}):
\begin{equation}
P(\Delta|p_\mathcal{B}^{(g)})\propto \frac{1}{M_2(\Delta|p_\mathcal{B}^{(g)},T_\gamma)}\exp\left(\frac{2}{r}\int^{\Delta}_0 d\Delta' \frac{M_1(\Delta'|p_\mathcal{B}^{(g)},T_\gamma)}{M_2(\Delta'|p_\mathcal{B}^{(g)},T_\gamma)}\right).
\label{stationarydistribution}
\end{equation}
As the notation emphasizes, the stationary probability distribution is dependent on the trading probabilities, and these are themselves function(al)s of the probability distributions $P(\Delta|p_\mathcal{B}^{(g)}),~g=1,2$. Accordingly we use the iterative procedure described above to find a solution, repeating steps (ii) and (iii) until the trading probabilities $T_\gamma$ remain stable to an accuracy of $10^{-6}$.

\begin{figure}[h!]
        \centering
			\includegraphics[width=\textwidth]{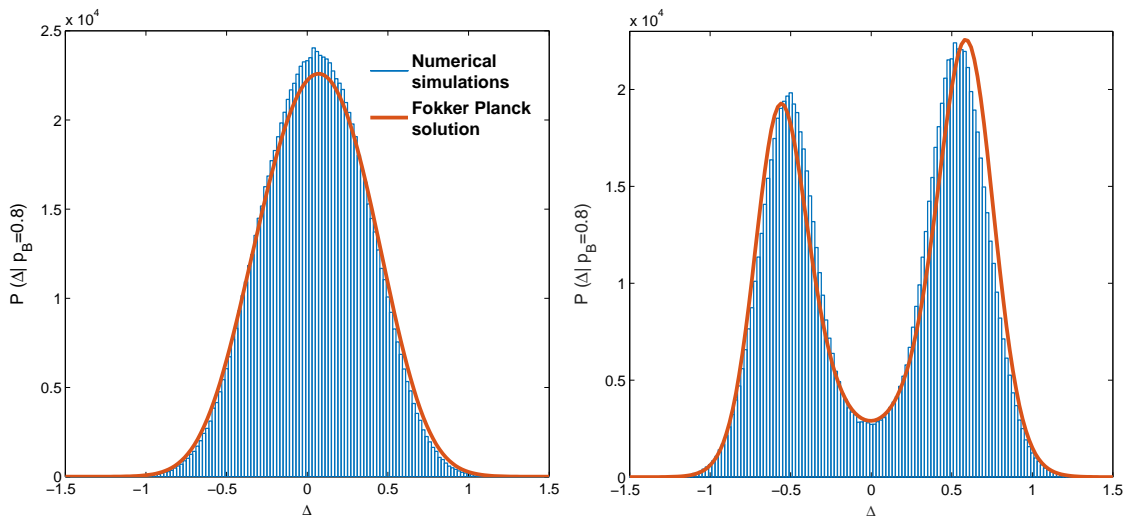}
         \caption{{\bf Steady states of the reduced model.} The distributions $P(\Delta|p_\mathcal{B}^{(1)}=0.8)$ from simulation and Fokker-Planck theory; the corresponding distributions for the second type of agents $P(\Delta|p_\mathcal{B}^{(2)}=0.2)$ are mirror images with respect to the y-axis.
Left: $\beta=1/0.45$, Right: $\beta=1/0.15$. Forgetting rate is $r=0.1$, other system parameters as specified in Table \ref{parametertable}.}
\label{fig4:ssbuyers}
\end{figure}

Fig.~\ref{fig4:ssbuyers} compares the steady state distributions $P(\Delta|p_\mathcal{B}^{(1)})$ obtained from direct numerical simulations and from the Fokker-Planck theory described above. The buying preferences $(p_\mathcal{B}^{(1)},p_\mathcal{B}^{(2)})=(0.8,0.2)$ are symmetric about 1/2 and so we do not show $P(\Delta | p_\mathcal{B}^{(2)})$, which would be the mirror image about $\Delta=0$.
The distribution obtained for low $\beta$ (Fig.~\ref{fig4:ssbuyers} left) shows the behaviour expected for an unsegregated population, with a single-peaked distribution of relative attraction $\Delta$ for each agent type; the mean of the distribution essentially coincides with the fixed point of the single agent dynamics, i.e.\ the solution of $M_1(\Delta)=0$. In the high-$\beta$ regime, the agents of each type segregate into two groups corresponding to the two peaks of the distribution of the attraction differences $\Delta$. As in the case of fully adaptive agents, one group is return-driven, i.e.\ prefers the market that awards them with higher returns, although they are not always able to trade there. The agents in the other group are volume-driven; they settle for the market where returns are lower on average but where they can trade more regularly.

The Fokker-Planck theory successfully reproduces the qualitative transition from unsegregated behaviour at low values of the intensity of choice to segregated steady states at higher values of $\beta$. The quantitative agreement with numerical simulations is good, remarkably so given that the latter were obtained for relatively small systems ($N=200$ agents) and for moderate $r=0.1$ while we developed the theory for the combined limits of large $N$ and small $r$. The agreement between theory and numerical experiment also suggests that  segregation is not a finite-size effect.

\paragraph{Characterising the phase transition.} To track the change in the shape of the relative attraction distributions from unimodal to bimodal as $\beta$ is changed we consider the so-called Binder cumulant (see for example ~\cite{Binder}). This quantity is defined as
\begin{equation}
B=1-\frac{\langle \Delta^4\rangle_{P(\Delta|p_\mathcal{B}^{(g)})}}{3\langle \Delta^2\rangle^2_{P(\Delta|p_\mathcal{B}^{(g)})}},
\label{binder}
\end{equation}
and is a good indicator of segregation as it has different limiting values for unimodal and bimodal distributions. In principle $B$ is dependent on the type of agent considered, $g$; but in the situations we consider where the two types have distributions of score differences $\Delta$ that are mirror images of each other, this dependence disappears. For numerical simulation data we show $(B^{(1)}+B^{(2)})/2$. The Binder cumulant takes the value $B=0$ for a Gaussian distribution, while $B\approx 2/3$ for a distribution consisting of two sharp peaks (the precise value depends on the relative weight of the peaks). For small $r$, where the $\Delta$-distributions become sharp around their peak(s) according to Equation (\ref{stationarydistribution}) these are therefore the values of $B$ we expect for $\beta$ below and above the segregation threshold $\beta_{\rm s}$, respectively. For finite values of the learning rate $r$ this will become a smooth transition between the two limiting values.
\begin{figure}[h!]
       \includegraphics[width=\textwidth]{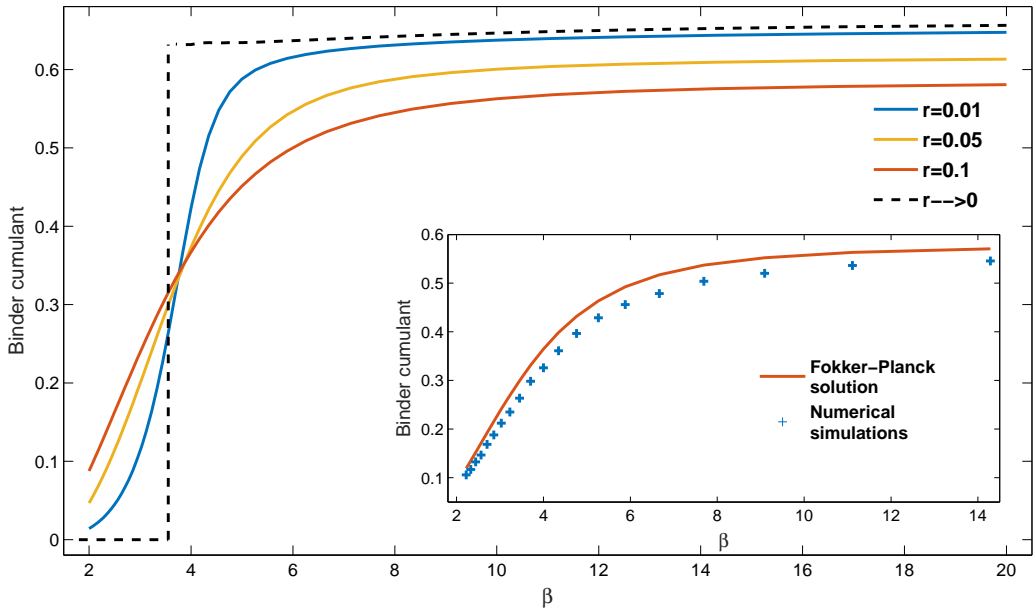}
       \caption{{\bf Binder cumulant} (as defined in Eq. (\ref{binder})) of the relative attraction distribution for different values of $r$ as shown in the legend (large to small $r$ from bottom to top on the right of the plot). Inset: Numerical simulations (blue) versus Fokker-Planck theory (orange) for $r=0.1$. Simulated system size $N=200$ traders, with $\Delta$ distributions obtained from the last $1/r$ trading periods of $100$ independent simulation runs.}
       \label{fig5:Binders}
\end{figure}

In Fig.~\ref{fig5:Binders} we show the predictions of our Fokker-Planck theory for the Binder cumulant of the reduced model with fixed buy/sell-preferences as a function of $\beta$, and for different values of $r$. These confirm the expectations set out above, with the segregation transition become increasingly sharp as $r$ decreases. The limit $r\rightarrow 0$ of the theory can be worked out by a separate procedure and gives a sharp transition at a well-defined segregation threshold $\beta_{\rm s}$ (for the system presented in Fig.~\ref{fig5:Binders}, $\beta_{\rm s}=3.55$). The inset of the Fig.~\ref{fig5:Binders} compares numerical simulations and theory for $r=0.1$ and again shows very good agreement. We attribute the remaining deviations to finite-size effects and to the fact that simulations necessarily operate at nonzero $r$ while the  Fokker-Planck theory is derived in the limit of small $r$.
The qualitative behaviour of the Binder cumulant, i.e.\ a transition between the expected theoretical values for small and large values of $\beta$ that becomes sharper with decreasing $r$, is exactly the same in the fully adaptive model; see~\cite{AEpublication}. There are of course quantitative differences; e.g.\ the threshold value $\beta_s\approx 5.9$ (for $r\rightarrow 0$) is somewhat higher~\cite{AEpublication}.

\paragraph{Further characterisation of segregation dynamics.} So far we have successfully constructed a mathematical description that reproduces the segregation effects seen in simulations. We now use the theory to look more closely at the emergence of segregation and the properties of the segregated state. To this end we consider average returns across the population of agents. These allow us detect whether segregation brings population-level benefits even though all agents make decisions on a purely individual basis. Given that persistence times are finite, the population-averaged returns also give the long-time average returns for any agent and so they tell us about the benefits of segregation for single agents.

\begin{figure}[h!]
       \includegraphics[width=\textwidth]{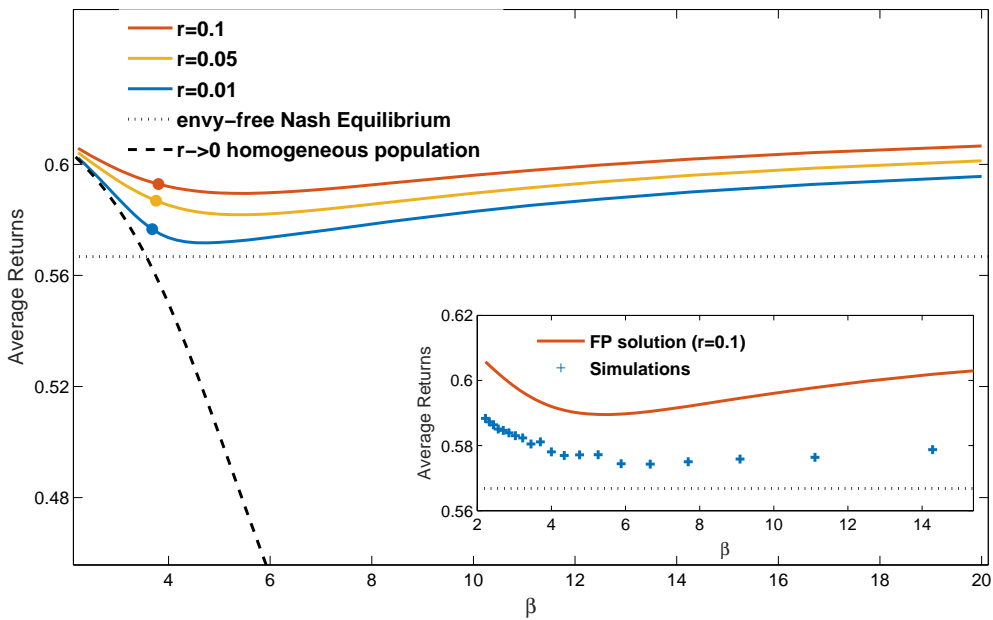}
       \caption{{\bf Returns.} Average steady state population returns against $\beta$ for different values of $r$ (see legend; $r$ decreases from top to bottom), as predicted by Fokker-Planck theory (circles denote $\beta_s$ for the given value of $r$). Black dashed line: average return of an unsegregated population for $r\rightarrow 0$. Horizontal dotted line: envy-free Nash equilibrium (see text). The segregated solution achieves higher returns than either the unsegregated population at the same $\beta$, or the Nash equilibrium. Insert: comparison between numerical simulations and Fokker-Planck theory for $r=0.1$ and population size $N=200$.}
       \label{fig6:Returns}
\end{figure}

In Fig.~\ref{fig6:Returns} we plot the average return obtained by agents in the steady state against intensity of choice $\beta$ for populations with different forgetting rates $r$. The data shown are from the Fokker-Planck analysis of the reduced two-strategy model described above. Remarkably, the average return is a non-monotonic function of $\beta$: it has a 
minimum close to the segregation threshold $\beta_{\rm s}$, at a level that decreases as $r$ is reduced. The qualitative non-monotonic trend is also found in numerical simulations (see inset). It is harder to detect there as the absolute changes in returns are fairly small, but appears in both the reduced and the fully adaptive model.

To put the returns for our segregated steady states into context, we compare them to two benchmark values. The dashed line in Fig.~\ref{fig6:Returns} indicates the first of these, which is the average return of an unsegregated population (in the limit $r\rightarrow 0$ limit, to be discussed in more detail shortly). While this homogeneous population return decreases monotonically with $\beta$, segregation avoids this decrease for intensities of choice $\beta>\beta_{\rm s}$ and in fact converts it to an increase.

As a second baseline we show in Fig.~\ref{fig6:Returns} the average return in an envy-free Nash equilibrium. A Nash equilibrium in general is a state in which no agent can increase their payoff by unilaterally changing strategy. In our system, a state would be specified by the probability $p^i_1$ for each agent $i$ to choose market 1 -- as the set of actions is $\{$Market 1, Market 2$\}$. The strategy of an agent is defined by $(p_1,1-p_1)$ -- and a Nash equilibrium is reached when no agent $i$ can achieve higher average return by changing $p^i_1$, at fixed strategies of all other players. This still leaves potentially many different Nash equilibria~\cite{Nicole} and so we focus on what we will refer to as the {\em envy-free} Nash equilibrium. This is an equilibrium in which no agent is envious of any other agent's return because they all achieve the same payoff on average over time. This is realized in our system when the average returns for all four distinct actions (buy or sell, at market 1 or 2) are the same. This condition allows us to identify a unique set of trading probabilities and from these the envy-free Nash equilibrium return. As  Fig.~\ref{fig6:Returns} shows, this is always {\em lower} than the average return of a segregated population (both in theory and simulations, see the inset). A segregated steady state is thus better in terms of returns than both the homogeneous steady state at the same $\beta$ and the envy-free Nash equilibrium. What is notable here is that the segregated state is not envy-free over short time horizons: an agent in the volume driven group obtains a lower return than one from the return driven group. However, as emphasized above, on very long time scales agents can change their loyalties, i.e.\ change group, so that in a long-time average all achieve the same return. This again emphasizes the co-operative nature of the segregated state.

\paragraph{Estimating the segregation threshold $\beta_{\rm s}$: the $r\rightarrow 0$ limit.} We now proceed to study the  segregation threshold $\beta_{\rm s}$ of the intensity of choices and how it depends on the parameters of the model. This is easiest in the limit $r\to 0$ where the segregation transition is sharp, see Fig.~\ref{fig5:Binders} above.
We first focus on the regime $\beta<\beta_s$, i.e.\ the unsegregated phase, in which the distributions $P(\Delta | p_{\mathcal{B}}^{(g)})$ will be single-peaked. In the limit  $r\to 0$, Equation (\ref{stationarydistribution}) indicates that (i) this peak becomes infinitely sharp, so that the distributions are of the form
$P(\Delta | p_{\mathcal{B}}^{(g)})=\delta(\Delta-\Delta^{(g)})$; and (ii) the location
 $\Delta^{(g)}$ of the peak for each group of players is determined by the zero-drift condition
\begin{equation}
M_1(\Delta^{(g)} | p_\mathcal{B}^{(g)}, T_\gamma)=0.
\label{zero_drift}
\end{equation}

The next step is to solve these two equations for $\Delta^{(1)}$ and $\Delta^{(2)}$, which is the $r\to 0$ analogue of finding the steady state of the FP equation for general $r>0$. As before one needs to make the solution self-consistent so that the trading probabilities $T_\gamma$ appearing in (\ref{zero_drift}) are those calculated from the distributions $P(\Delta | p_{\mathcal{B}}^{(g)})=\delta(\Delta-\Delta^{(g)})$ themselves. The iterative approach explained above can again be used to find such a self-consistent solution, starting from $(\Delta^{(1)},\Delta^{(2)})=(0,0)$.

We briefly explore the properties of this homogeneous steady state before considering how to detect the onset of segregation. We focus on agent types with symmetric biases toward buying and selling, $p_\mathcal{B}^{(1)}=1-p_\mathcal{B}^{(2)}$. For this choice we find non-zero $(\Delta^{(1)},\Delta^{(2)})$ even in the limit $\beta\to 0$. This indicates that agents recognize the more rewarding option: agents that are more likely to buy have a preference for the market that is good for buyers, and similarly for sellers. Of course $\beta\to 0$ means that agents nevertheless choose randomly between the markets.
As $\beta$ increases, the relative attractions $(\Delta^{(1)},\Delta^{(2)})$ become more pronounced, i.e.\ they move away from zero, and agents start to choose the ``better'' market more frequently. This is the reason for the decay of the average return in the homogeneous steady state with $\beta$, as shown in Fig.~\ref{fig6:Returns}: as agents of each type increasingly focus on ``their'' market, buyers congregate in one market and sellers in the other; trading opportunities are reduced and thus the average return (which includes zero returns for trading periods where an agent cannot trade) decreases. The more definitive choices agents make are the consequence of the two effects:  (1) the fixed points $(\Delta^{(1)},\Delta^{(2)})$ increase in absolute value (and they correspond to the difference in the average score an agent receives in the two markets); (2) the increase of $\beta$ makes the choices more definitive. Quantitatively, we find that (2) is the stronger effect.
As the intensity of choice $\beta$ is increased further, we expect a transition to a segregated steady state. To detect this transition we can follow the general logic explained in the beginning of Analytical Description: a segregated state must have more than one fixed point of the single-agent dynamics. The peak position $\Delta^{(1)}$ is always a fixed point for agent type 1 from Eq. (\ref{zero_drift}), and similarly for type 2. To detect the onset of segregation we therefore need to check when {\em additional} fixed points appear, i.e.\ additional solutions of the zero-drift condition $M_1(\Delta | p_{\mathcal{B}}^{(g)},T_\gamma)=0$. In looking for these alternative fixed points we need to keep the trading probabilities $T_\gamma$ {\em fixed} at their values calculated for the homogeneous steady state, because we are considering the {\em single-agent} fixed points. Note that in general one needs to search for alternative fixed points globally, i.e.\ across all possible $\Delta$. This is because for most parameter settings the new fixed points appear far from $\Delta^{(1)}$ or $\Delta^{(2)}$, respectively, as pairs of stable and unstable fixed points.

\begin{figure}[h!]
	   \includegraphics[width=\textwidth]{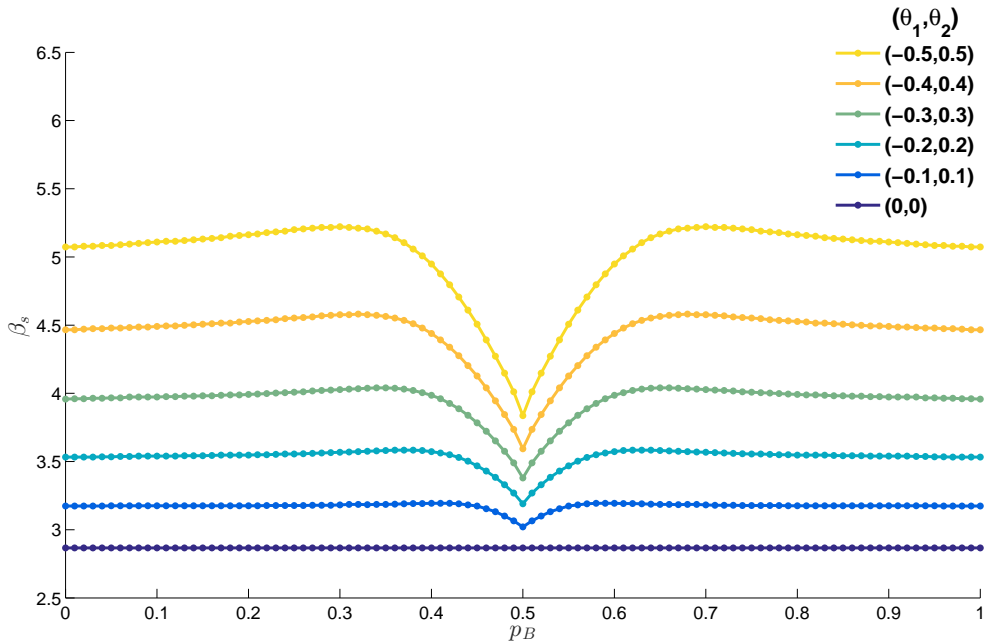}
       \caption{{\bf Segregation thresholds for different symmetric agent types, at different markets.} Segregation threshold as a function of the first agent type's preference for buying $p_\mathcal{B}=p_\mathcal{B}^{(1)}$, assuming the other agent type has the opposite preference $p_\mathcal{B}^{(2)}=1-p_\mathcal{B}$. Different sets of symmetric market parameters are compared; non-trivially, the region of segregation is largest when both markets are fair.}
       \label{fig7:pbDiffThTcsurf}
\end{figure}

\begin{figure}[h!]
       \includegraphics[width=\textwidth]{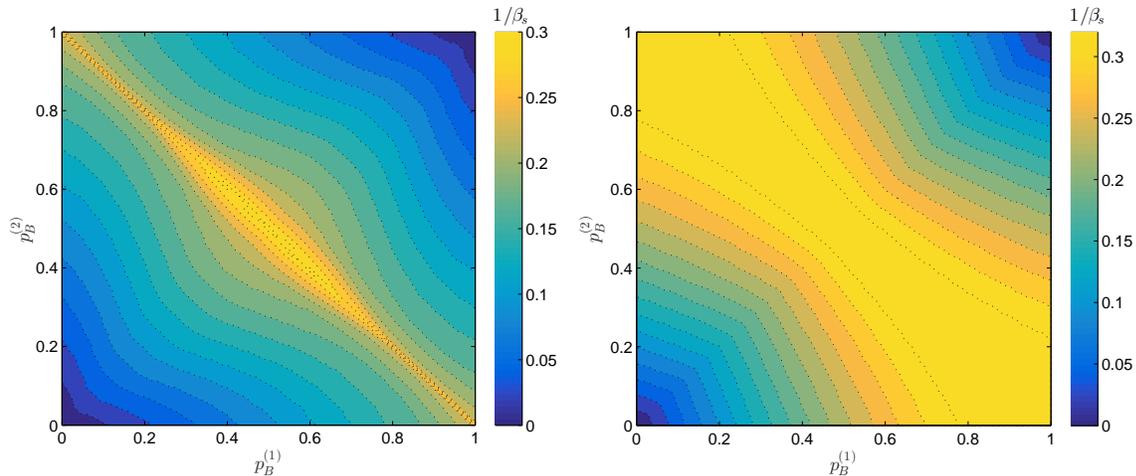}
       \caption{{\bf Segregation threshold for various agent types.} Contour plots of the segregation threshold $\beta_{\rm s}$ when preferences for buying $p_\mathcal{B}^{(1)}$ and $p_\mathcal{B}^{(2)}$ of both agent types are varied independently. Left: symmetric markets $(\theta_1,\theta_2)=(-0.2, 0.2)$; Right: two fair markets ($\theta_1=\theta_2=0$). Contours are presented in terms of $1/\beta$ for visual clarity.}
       \label{fig8:pb1pbTcsurf}
\end{figure}

In Figs.~\ref{fig7:pbDiffThTcsurf} and  \ref{fig8:pb1pbTcsurf} we present the segregation thresholds obtained by the method above, for various parameter settings. In Fig.~\ref{fig7:pbDiffThTcsurf} specifically we show how $\beta_{\rm s}$ changes with $p_\mathcal{B}$ for various symmetric markets. One observes that $\beta_{\rm s}$ does not depend very strongly on the exact preferences for buying of the agents, except for the region of parameters where both agent types have almost even preferences for buying and selling. A full analysis of the change in monotonicity in that region exceeds the scope of this publication and will be presented elsewhere. In contrast, the effect of the market biases follows a simple trend: the region of segregation shrinks as the difference between the two markets increases, suggesting that segregation is not a trivial effect of market biases.

In the Fig.~\ref{fig8:pb1pbTcsurf} we show two contour plots to compare the trends in $\beta_{\rm s}$ when the two agent types' buying preferences are varied independently, for two different choices of market parameters. On the left is the symmetric markets setup $(\theta_1,\theta_2)=(-0.2,0.2)$ that we have already used several times. From the plot we observe that the segregation threshold $\beta_{\rm s}$ is lowest when the subgroups are symmetric with respect to buy-sell preferences, i.e.\ $p_\mathcal{B}^{(1)}+p_\mathcal{B}^{(2)}=1$; the variation of  $\beta_{\rm s}$ along that line is presented in Fig.~\ref{fig7:pbDiffThTcsurf}. We also note that when the two agent types have similar preferences (e.g.\ both prefer buying over selling) then $\beta_{\rm s}$ is on average higher than in the case where the types have opposite preferences. This qualitative behavior we see also in the case of two fair markets that set the trading price at the equilibrium price, as shown in Fig.~\ref{fig8:pb1pbTcsurf} (right). We do not show similar plots for asymmetrically biased markets; the qualitative behaviour is similar there, but the line of minimal $\beta_s$ is no longer $p_\mathcal{B}^{(1)}+p_\mathcal{B}^{(2)}=1$ but $p_\mathcal{B}^{(1)}+p_\mathcal{B}^{(2)}=c$, where $c$ is a constant that depends on the market parameters.

Finally, we can adapt the above method of calculating the segregation threshold to the original model of fully adaptive agents choosing among all four possible actions (e.g. $\gamma \in \{\mathcal{B}1,\mathcal{S}1,\mathcal{B}2,\mathcal{S}2\}$). Because of the difficulties of finding the steady state solution numerically for finite $r$, this analysis is carried out only in the limit $r\to 0$. Here it is feasible because we only need to find zeros of the drift $\mathbf{M}_1$, rather than solve for a full distribution $P(\mathbf{A})$ that is a stationary solution of the Fokker-Planck equation. As argued before, the first moment ($\mathbf{M}_1$) is a function of the whole distribution, but in the $r\rightarrow 0$ limit and for $\beta < \beta_s$ this distribution is a delta distribution, which simplifies the calculation.

\begin{figure}[h!]
	   \includegraphics[width=0.5\textwidth]{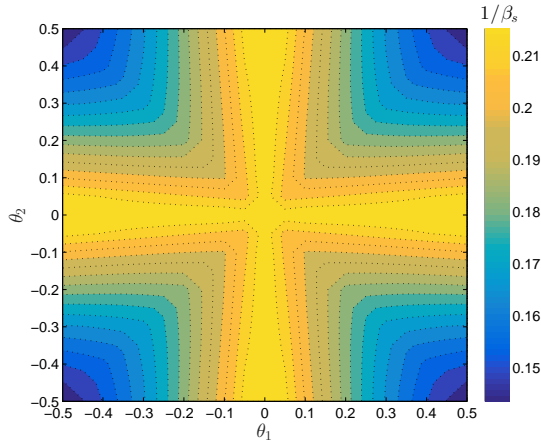}
       \caption{{\bf Segregation threshold for fully adaptive traders.} Contour plots of the segregation threshold $\beta_{\rm s}$ in a system of fully adaptive agents, against the market bias parameters $(\theta_1, \theta_2)$. Along the lines where at least one market is fair, $\beta_{\rm s}$ is lowest so that segregation occurs for the widest possible range of intensities of choice $\beta$.}
       \label{fig9:adaptiveTcsurf}
\end{figure}

In Fig.~\ref{fig9:adaptiveTcsurf} we show a contour plot of the segregation threshold against the two market biases. As in Fig.~\ref{fig7:pbDiffThTcsurf}, for agents with fixed buy-sell preferences, we notice that the segregation threshold $\beta_s$ decreases when the difference in the symmetric market biases decreases; this can be seen specifically by looking at the symmetric markets diagonal of Figs.~\ref{fig9:adaptiveTcsurf} and ~\ref{fig7:pbDiffThTcsurf}.  Additionally in the system of fully adaptive agents we notice that this conclusion extends to the case when at least one market is fair, i.e.\ segregation then also arises for smaller values of $\beta$. 
In Fig.~\ref{fig9:adaptiveTcsurf} we note also a four fold symmetry, which we conjecture is a consequence of the fact that agents have more choices: a fully\ adaptive agent does not have a preferred market initially, so all four action choices are equal. This is not the case for an agent with a fixed preference for buying and selling because returns are buyer/seller-specific. The four fold symmetry in the Fig.~\ref{fig9:adaptiveTcsurf} also tells us that the absolute value of $\theta$ is enough to describe the market. This further means that the commonly investigated choice of two symmetric markets is analogous to the case of two identical markets and we note that the segregation threshold is lowest when both markets are fair. This is in agreement with the results of the model with agents with fixed buy-sell preferences, see specifically in Fig.~\ref{fig8:pb1pbTcsurf} where for every choice of $(p_{\mathcal{B}}^{(2)},p_{\mathcal{B}}^{(2)})$ the segregation threshold $\beta_s$ is smaller when the markets are fair.

\paragraph{Multiple steady states.} When iteratively finding a self-consistent steady state of the Fokker-Planck equation as discussed around Equation (\ref{AdaptiveFokkerPlanck}), we always used as initial condition a distribution of all agents having zero attractions. While the dynamics of the iterative solution is not identical to the real dynamics, this choice was sensible to obtain steady states that match the ones from simulations of the real dynamics as closely as possible. After further analysis we found that in the numerical simulations for small values of $r$ one can obtain two qualitatively different classes of segregated states. To investigate this notion further, we now explore whether there can be more than one self-consistent solution of the Fokker-Planck equation. Such additional solutions could be accessible for example by using our iterative procedure, but starting from other initial conditions.

The simplest way to perform this exploration systematically is to realize that the trading probabilities $T_\gamma$ from Equation (\ref{tradingP}) only depend on the demand-to-supply ratios $D_1$ and $D_2$ at the two markets; we define $D_m$ for each market $m$ as the ratio of the number of buy and sell orders that arrive at the market. Specifying $(D_1,D_2)$ thus tells us all trading probabilities, and hence determines a unique steady state solution of the Fokker-Planck equation. From this steady state solution we can recalculate $D_1$ and $D_2$, and plot in the $(D_1,D_2)$-plane the two lines where the new and old $D_1$ ($D_2$, respectively) coincide. The intersections of these two lines are then the self-consistent steady states we are after. We find numerically, in the range of parameter values that we have explored, that there is either one such state or there are three.

\begin{figure}[h!]
\center
\includegraphics[width=\textwidth]{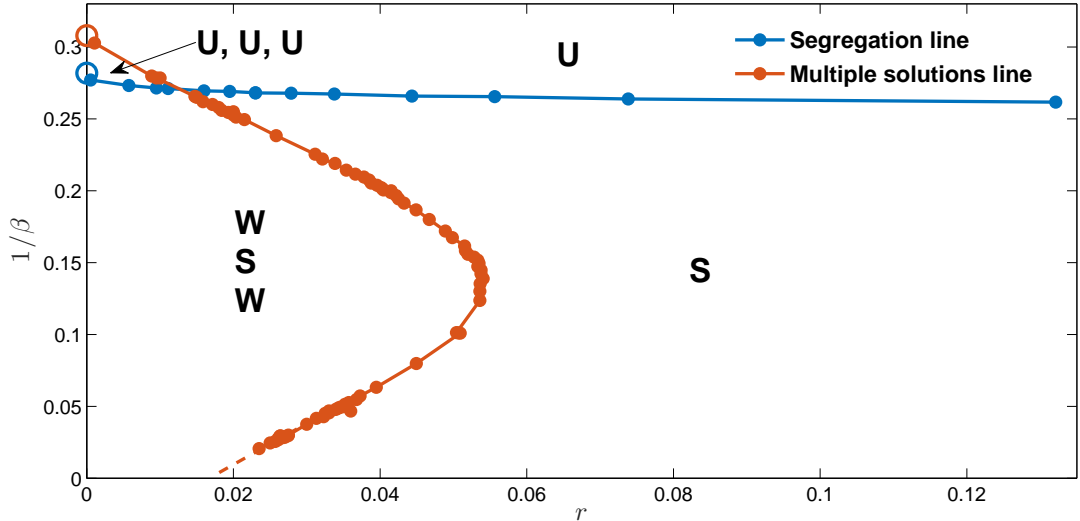}
\caption{{\bf Regions of different steady states in the space of parameters $r$ and $\beta$.} Number and nature of the steady state solutions of the Fokker-Planck equations in the $(r,1/\beta)$-plane (U: unsegregated state, S: strongly segregated state, W: weakly segregated state). Other system parameters are set to their default values (see Table \ref{parametertable}). To the left (small $r$) of the orange line there are three steady state solutions, while to the right there is only one. The blue line separates the region where at least one of the solutions is segregated, i.e.\ bimodal. Empty circles represent $r\to 0$ endpoints of the two lines, calculated independently. }
\label{fig10:rtphasespace}
\end{figure}

In Fig.~\ref{fig10:rtphasespace} we show the resulting phase diagram in the $(r,\beta)$-plane and indicate the number and nature of self-consistent steady state solutions. Where solutions are segregated we differentiate between two possibilities, strong and weak segregation. By strong segregation we mean a solution branch where the two peaks in $P(\Delta | p_\mathcal{B})$ remain of comparable height as $r\to 0$. We call a solution branch weakly segregated when one peak weight becomes exponentially small for $r\to 0$, so that the solution effectively reduces to an unsegregated one. In the limit, all agents then typically prefer the same market.
The empty circles in the figure are results from a separate analysis of the deterministic ($r\rightarrow 0$) theory, and are consistent with the extrapolation of the results for nonzero $r$. Numerical simulations confirm the existence of all three solution types (U, S, W) in the relevant regions of the phase space, where necessary starting from appropriately tuned initial conditions.
Where the strongly segregated solution competes with others it appears to become unstable but long lived in small system simulations. (Details are beyond the scope of this paper and will be given elsewhere.)

The blue line in Fig.~\ref{fig10:rtphasespace} is the segregation threshold $\beta_{\rm s}$ discussed previously. It is nearly constant, increasing only very slowly with $r$; this shows that the $\beta_{\rm s}$ values obtained via the $r\to 0$ analysis will be good estimates also for larger $r$. A second interesting feature of Fig.~\ref{fig10:rtphasespace} is that there is a threshold value of $r\approx 0.054$ above which the strongly segregated solution is the only possible option, for any $\beta$ large enough for a segregated state to exist at all: this state is then the genuine steady state. Where agents have only moderate memory, with $1/r$ of the order of 10 trading periods, steady states where one market comes to attract most agents thus disappear, and the emergence of persistent market loyalties becomes the norm.
For the fully adaptive model we have preliminary evidence for a similar phase diagram as in Fig.~\ref{fig10:rtphasespace} and in particular for the existence of weakly segregated states, but defer further analysis of their behaviour to a separate publication.

\section*{Discussion}

The increasing fraction of global trades that now takes place online in the form of high-frequency algorithmic trading (around $30\%$ of equity trading in the UK and around $60\%$ in the USA~\cite{foresight}) raises many challenges for market regulations. To make such regulation as informed as possible, research is needed to understand the possible long-run states of such systems. We have focused on one particular long-run state -- the segregation of traders. As argued in the introduction, it is easy to imagine that such a state can bring benefits to the system, but also disadvantages; both make it important to investigate and understand the phenomenon and its consequences. We have developed a stylized model of double-auction markets serving a large number of adaptive traders; this has helped us to shed light on the benefits of segregation for the population as a whole.

We have considered two variants of our model, one populated with agents who can adaptively tune their preferences for buying and selling along with their preferences for the two markets; in the other model agents have fixed preferences for buying and hence also selling. The two models share our main qualitative result: above a threshold value $\beta_{\rm s}$ of the intensity of choice $\beta$ the agents segregate, i.e.\ develop a long lasting loyalty to one action, or one market. The onset of segregation is signalled both in simulations and in the analytical description by the emergence of multiple peaks in the distribution of agent preferences, as summarized e.g.\ in the Binder cumulant. These peaks are accompanied by long persistence times for the agents to remain within each peak~\cite{AEpublication}.

In the model with fixed buying preferences we have shown that one can develop an analytical description of segregation to a significant level of detail. We find that even though individual agents do not explicitly try to maximise the well-being of the entire population, the strongly segregated state is in effect cooperative: it is more beneficial for every individual player, and the population as a whole, compared to possible unsegregated and envy-free Nash equilibrium states. The segregated state is neither envy-free -- the traders that specialize to what we called volume-driven behaviour have lower returns in the short term than the return-driven ones -- nor a Nash equilibrium. In this sense the segregated state is stabilized by incomplete information, of each agent about the precise returns to be expected from each action and about the average returns achieved by others.

The transition from the homogeneous to the segregated state is caused by increasing the intensity of choice parameter $\beta$. There are two ways of interpreting this. One is that as $\beta$ grows, agents optimize against return differences on smaller and smaller scales $1/\beta$: our results then show that the more stringently agents optimize their behaviour, the higher the likelihood of segregation. Alternatively, as $\beta$ affects the agents' preferences only via products with the attractions, which themselves are proportional to the returns, an increase in $\beta$ has the same effect as an increase in the scale of returns at fixed $\beta$. When the possible returns are small an agent then plays randomly, while if the stakes are high an agent will try to take into account information from previous trades as much as possible. In this interpretation our main result states that there is a critical scale for single transaction returns above which the preferred state of the agent population is the segregated one.

We studied in some detail the dependence of the segregation threshold $\beta_{\rm s}$ on model parameters. One intriguing finding is that the threshold is generally lowest when the two markets are similar, demonstrating that segregation is not trivially driven by market differences. The precise value of $\beta_{\rm s}$ is determined by collective behaviour, with agents continually adjusting to trading conditions the population itself creates. This rules out simple intuitive estimates of $\beta_{\rm s}$. It is reassuring, however, that the quantitative variations in $\beta_{\rm s}$ are small for the different models we investigated, with values lying in the range $0.1\ldots0.3$ across most of the parameter space of our models.

Finally, it is important to discuss simplifications in our analysis and the way they might have affected our main results. We have made a number of assumptions about the bid and the ask distributions. It turns out that when these are relaxed, our main results change only quantitatively but not qualitatively. For example, we have shown all data for the case where the mean bid is greater than the mean ask, which enables more trades at the markets. However, qualitatively identical results are obtained in the opposite case, with the only change being in the specific value of the threshold $\beta_{\rm s}$. The specific assumptions we made on the shape of bid/ask distribution can also be relaxed. In fact, within the Fokker-Planck description only the first and the second moment of the truncated bid/ask distribution appear and the precise shape of the distribution is otherwise immaterial.

Another important assumption about the agents is their learning rule. As we discussed when first setting out the model, in the literature there are several variants of the rule we used. There is broad agreement on the existence of an intensity of choice-like parameter in the learning dynamics, see e.g.~\cite{EWA,EWAnew,luckystar,CrutchfieldSato2003,GallaFarmer2013,Cheung97,Goeree} for applications and experimental confirmations of this. As regards score updating, in many cases agents are assumed to update the score for unplayed actions with the corresponding fictitious score. We have argued that this is not realistic in our set up; however, changing the way an agent forgets the score of the unplayed actions might be reasonable. This suggests a generalized reinforcement rule that is still in the class of EWA \cite{EWA,EWAnew} approaches: 
\begin{equation}
A_{\gamma}(n+1) = \begin{cases}
(1-r)A_{\gamma}(n) + rS_{\gamma}(n), & \mbox{if the agent chose action $\gamma$ in round $n$}\\
(1-\alpha r)A_{\gamma}(n), & \mbox{if the agent chose an action $\delta\neq\gamma$ in round $n$}
\end{cases}
\end{equation}
Here the parameter $\alpha$ can be tuned between two extremes: $\alpha=1$ is the case we have discussed so far, where an agent forgets the score of the unplayed actions with the same rate as the scores of the played action, effectively imputing a score of 0 for actions that were not taken. On the other hand, $\alpha=0$  corresponds to the case where an agent does not forget the attractions of the unplayed actions, which effectively means imputing a score equal to the average score up to that time. Without going into the details, our analytical methods can be extended to cover the entire range of $\alpha$ for both variants of our model (fully adaptive and fixed buy-sell preference). We find generally that the segregation threshold $\beta_{\rm s}$ increases as $\alpha$ decreases towards zero. It remains finite for any $\alpha>0$, so that segregation continues to take place, in a manner qualitatively similar to our vanilla model ($\alpha=1$). The case $\alpha=0$, which presupposes that agents have effectively infinite memory to scores of unplayed action, is special and will be explored further elsewhere.

We have also made the simplifying assumption that budget constraints on the agents can be ignored. Above the segregation threshold our results show substantial persistence times for agents in a particular role (e.g.\ persistent buying, at one of the two markets). While this may be in apparent conflict with budget or stock constraints, it is worth remembering that agents do change their loyalty eventually so they just need a large enough budget to sustain a long period of buying that is then followed by a long period of selling. Also, while persistence times do get exponentially large for very small $r$, our results show (cf.~Fig.~\ref{fig10:rtphasespace}) that segregation can occur up to fairly large values of $r$. In this regime persistence times -- while longer than in homogeneous states -- are only moderately large so that budget constraints should be relatively easy to satisfy. We also note that in T\'oth et al.~\cite{Toth2015}  the authors show that the well documented persistence in orders of the same sign (i.e. an order to buy tends to be followed by more orders to buy and similar for an order to sell) at the shorter time scales is dominated by a single trader splitting his/her order; also the tendency to buy or sell persistently was shown to be stronger than collective effects such as order herding. Finally, segregation also occurs in models with explicit budget constraints, as discussed in the Outlook below.

One question that remains, and which is not easy to address in full generality, is whether the emergence of segregation is an intrinsic property of systems with adaptive agents, or a consequence of our specific stylized model. The simplicity of the model itself argues for the former, as we did not need to make exotic assumptions to find segregation. But clearly there is still much to do from here to reach a detailed understanding of segregation in real markets, at a level that can directly influence policies. An initial step in this direction is described in the following section. 

\begin{figure}[h!]
\center
\includegraphics[width=\textwidth]{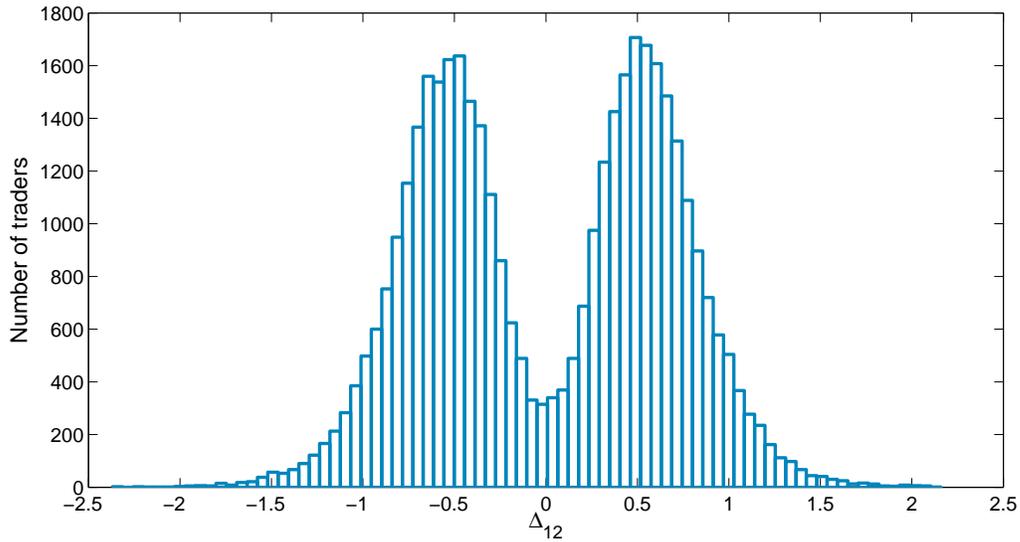}
\caption{{\bf Distribution of attractions to markets in a continuous double auction setting.} We show distributions $P(\Delta_{12})$ after $100$ trading periods, each consisting of an initial call to populate the limit order book and then $2000$ trading steps in which a random agent is selected to consider the limit order book and decide whether to execute an existing order or submit a new one. The distribution is an average over $10$ independent runs of a system with $N=2000$ traders, whose learning parameters are $1/\beta=0.07$ and $r=0.05$.}
\label{fig11:contmarket}
\end{figure}

\paragraph{Outlook.} Our work was motivated by the presence of segregation noticed in CAT tournaments \cite{CAToverview,CATcosegregation} where both the trading strategies and market mechanisms were far more complex and realistic. To understand how segregation can emerge we did not analyse a system populated with the same complex parts, but instead hypothesized that segregation is a consequence of mutual co-adaptation of traders and markets. Having established that segregation arises generically when agents adapt their preferences according to a well studied reinforcement learning scheme, an obvious direction for future work is to investigate whether removing various restrictions in our modelling approach might affect the occurrence of segregation. As an initial signpost along this route we report briefly on results we have obtained for a well-studied model of markets and traders~\cite{HuberJcurve, Huber2008, TothJcurve,DaanJcurve} that has more realistic assumptions than ours in a number of respects (see Methods section for details). The market is a continuous double auction with an open limit order book, the agents have budget constraints and trading strategies take account of fundamentals like dividend returns. We extend this model minimally by assuming there are two markets rather than one, and by allowing traders to choose between markets adaptively using the reinforcement rule Eq.~(\ref{reinfrule}) with a natural definition of the relevant score. In Fig.~\ref{fig11:contmarket} we show the resulting distribution of attractions towards each of the two markets after $100$ trading periods. The key observation is that segregation still emerges, in a model that differs from ours in having (1) budget constraints, (2) complex trading strategies and (3) continuous double auctions. This supports our view that segregation should be relatively generic when agents can choose between multiple markets and do so adaptively. 

In other future work it will be interesting to link to work~\cite{Cheung97} that draws on evidence from Behavioural Game Theory and suggests that values of $\beta$ are consistent across games but increase in more informative environments. The authors argue that a parameter closely analogous to $r$ also increases with the trustworthiness of information in the system. Bearing in mind the results shown in Fig.~\ref{fig10:rtphasespace}, where for large $r$ and large $\beta$ the only steady state is the segregated one, this suggests that more informative environments, or ones where information is more trustworthy because e.g.\ of stability over long timescales, might naturally lead to segregated states. It would be exciting to study this effect explicitly in a suitably extended model, or possibly in experiments with market games. Further directions for investigation might include agents with heterogeneous learning parameters \cite{Nicole}, a feature that will clearly be present in markets that are used both by individual investors and e.g.\ large funds. Additionally, the markets themselves could be made adaptive, with the aim of understanding whether the resulting competition between markets will suppress or promote segregation. There is clearly also scope for capturing more complex trading strategies in models like ours, something we have postponed as we were motivated by the simplicity of high-frequency trading algorithms and more broadly by the aim to develop a baseline model on which further extensions can be built.

\section*{Methods}
\subsection*{Numerical Simulations}

The parameters of the system are summarized in Table \ref{parametertable} with their definitions and the standard values that are used in the main text unless otherwise specified. For all numerical simulations (adaptive) agents were initialized with \ $A_\gamma=0~ \forall \gamma \in\{\mathcal{B}1,\mathcal{S}1,\mathcal{B}2,\mathcal{S}2\}$ and the trading dynamics was run until a stationary state was reached. Similarly in the simplified model with fixed preferences for buying, $A^{(g)}_{1,2}=0$ for all agents initially. The time required to reach a steady state was highly dependent on the chosen parameters (longer for low $r$ and high $\beta$), but for the results presented in this paper was in the range between $1,000$ and $10,000$ trading periods. To be sure that the system reached the steady state we measured average returns, attractions and higher cumulants of the attraction distributions; when the averages of these observables became independent of time we assumed stationarity and started collecting the data for the analysis. Statistics presented in the paper are gathered for each parameter setting from $100$ independent runs of the stochastic dynamics. Time averages are usually taken over $10/r$ trading periods at the end of each simulation run.

\paragraph{Continuous double auction markets.} For the preliminary results presented in the Outlook, we consider a system of market and traders introduced by T\'oth et al.~\cite{TothJcurve}. The market is a continuous double auction market with an open order book. Each trading period consists of an open call, during which every trader submits an order to populate the limit order book, followed by rounds during which a random trader can inspect the limit order book and then decides whether to execute an already existing order or to submit a new one. This decision process is governed by a fundamentalist trading strategy as follows. In the model, there is an underlying dividend process that determines the value of a stock. Different traders have different numbers of future dividend values accessible to them, starting from no information (thus, close to our Zero Intelligence traders) to information about $L$ future dividend values. (In previous studies $L$ ranged from $3$ to $10$; we use $L=4$, giving a total of five different types of agents including the uninformed ones). Based on their knowledge of future dividends, agents evaluate their stock holding using a dividend discount model, e.g.\ Gordon's growth model\cite{Gordon62}. This valuation is used as a private value based on which an agent decides whether to buy (when the best ask is lower than his/her private value), sell (when the best bid is higher than the private value) or submit a new order. These details of the trading strategy are the same as in~\cite{TothJcurve}; generally and unless mentioned otherwise, we follow exactly the existing model. The private value of an uninformed trader is a normally distributed random variable centred at the most recent trading price. Agents also have initial endowments in terms of number of stocks and available cash: in previous studies each agent had $40$ shares and the equivalent value in cash to start with, but as we look at much longer simulations, of $100$ to $1000$ trading periods as opposed to $10$--$30$, we increase the initial wealth to $100$ shares plus the equivalent value in cash.

When we introduce the second market, agents choose not only whether to buy or sell, based on dividend information as above, but also where they will trade; we assume they use the reinforcement rule Eq.~(\ref{reinfrule}). As before this relies on scores the agents assign after each trade. Because of the existence of a limit order book we need to differentiate between the Aggressor (the trader who executes an order from the limit order book) and the Quoter (the trader whose order was waiting in the limit order book) in assigning scores. As the trade happens at the price of the order existing in the limit order book, then for the {\em aggressor} we have his/her private value and the trading price, so the scores are as before: $S(t)={\rm pv}(t)-\pi(t)$ if the agent buys (as the s/he values the stock more) and $S(t)=\pi(t)-{\rm pv}(t)$ if the agent sells (as s/he believes the stock is worth less). On the other hand, the {\em quoter}'s order was waiting in the limit order book, so his/her private value may have changed from the moment the order was submitted. We therefore calculate the returns as $S(t)=\pi(t-1)-\pi(t)$ if the quoter buys and $S(t)=\pi(t)-\pi(t-1)$ if s/he sells: buyers value a price decrease while sellers value a price increase. We ran large systems ($N=2000$ as opposed to $N=100$ in the original works) and took the number of trading rounds within a period as $N$, so that on average every trader is chosen once to observe the order book and make a trading decision.

\begin{table}
\caption{
{\bf List of parameters defining the system; unless otherwise stated, the typical values are used.}}
\begin{tabular}{llr}
\multicolumn{3}{c}{{\bf System definition in terms of parameters}} \\
\textit{Parameter}   & \textit{Description} & \textit{Typical Value}  \\
$M$      & Number of markets    & $2$     \\
$(\theta_1,\theta_2)$ & Market biases - usually assumed to be symmetric, $\theta_1=-\theta_2$ & $(-0.2,0.2)$ \\
$N$      & Number of traders    & $200$     \\
$(\mu_a,\sigma_a)$ & Mean and standard deviation of the ask distribution (sellers) & $(9.5, 1)$\\
$(\mu_b,\sigma_b)$ & Mean and standard deviation of the bid distribution (buyers) & $(10.5, 1)$\\
$r$ & Forgetting rate; range $r\in[0,1]$ & $r=0.1$ \\
$\beta$ & Intensity of choice; simulation range $\beta \in [1, 50]$ & --\\
$(p_\mathcal{B}^{(1)},p_\mathcal{B}^{(2)})$ & Preferences for buying of the two types of agents of the toy model & (0.8,0.2)\\
\end{tabular}
\label{parametertable}
\end{table}

\section*{Acknowledgments}
We acknowledge funding by the Engineering and Physical Sciences Council EPSRC (UK), grant reference EP/K000632/1  (Network Plus, Towards consensus on a unifying treatment of emergence and systems far from equilibrium). PS acknowledges the stimulating research environment provided by the EPSRC Centre for Doctoral Training in Cross-Disciplinary Approaches to Non-Equilibrium Systems (CANES, EP/L015854/1). We thank Yuzuru Sato for useful discussions.

\section*{Author contributions}
Conceived and designed the experiments: A.A., P.S., P.M. Performed the experiments: A. A. Analyzed the data: A. A. Contributed reagents/materials/analysis tools: A. A., P. S., T. G. Wrote the paper: A. A., P. S., P. M., T. G.

\end{document}